\newcommand{\keyword}[1]{\index{#1}#1}
\def\onlinecite{\cite}

\def\Bm{{\rm \bf B}}
\def\Em{{\rm \bf E}}
\def\E{{\cal E}}
\def\Hm{{\rm \bf H}}
\def\H{{\cal H}}
\def\MC{Monte Carlo}
\def\Nc{N_{\rm c}}
\def\Nm{{\rm \bf N}}
\def\Sm{{\rm \bf S}}
\def\Um{{\rm \bf U}}
\def\Vm{{\rm \bf V}}
\def\betah{\hat\beta}
\def\br#1{\langle {#1}}
\def\bra#1{\langle {#1} |}
\def\cfMC{\keyword{correlation function \MC}}
\def\cf{{\it cf.}}
\def\d #1 #2{d_{#1}^{\,(#2)}}
\def\dMC{diffusion \MC}
\def\dh #1 #2{\hat{d}_{#1}^{\,(#2)}}
\def\etal{{\it et al.}}
\def\fh{{\hat f}}

\def\ket#1{| {#1} \rangle}
\def\psiTh{{\hat\psi}}
\def\psiT{{\tilde\psi}}
\def\psig{\psi_{\rm g}}
\def\qMC{quantum \MC}
\def\qmechal{\qmech al}
\def\qmech{quantum mechanic}
\def\rh{{\hat r}}
\def\r{{\bf r}}
\def\tE{\tilde E}
\def\tpsi #1{\tilde{\psi}^{(#1)}}
\def\tr #1{#1^T}

\def\prl{Phys.\ Rev.\ Lett.\ }
\def\pra{Phys.\ Rev.\ A }

\def\pre{Phys.\ Rev.\ E }
\def\jcp{J.\ Chem.\ Phys.\ }

\def\lesssim{\mathrel{\mathpalette\vereq<}}
\def\alt{\lesssim}
\def\vereq#1#2{\lower3pt\vbox{\baselineskip1.5pt \lineskip1.5pt
\ialign{$\hfill##\hfil$\crcr#2\crcr\sim\crcr}}}

\documentstyle[sprocl]{article}
\makeindex
\begin{document}
\title{Trial function optimization for excited states of van der Waals
  clusters} 
\author{M. P. Nightingale and Vilen Melik-Alaverdian}
\address{ Department of Physics, University of Rhode Island, Kingston
  RI 02881, USA} 
\maketitle

\abstracts{ A method is introduced to optimize excited state trial
  wavefunctions.  The method is applied to ground and vibrationally
  excited states of bosonic van der Waals clusters of upto seven
  particles.  Employing optimized trial wave functions with three-body
  correlations, we use correlation function \MC\ to estimate the
  corresponding excited state energies.  }

\section{Introduction}
\label{section.intro}

Solving the Schr\"odinger equation is a fundamental problem, but also
it is a timely one in view of the rapid experimental progress made in
recent years.\,\cite{ToenniesVilesovWhaley} For example, spectra of
van der Waals clusters with embedded chromophores have been measured
and have been used to construct more accurate interatomic pair
potentials and to test proposed three-body potentials.  The ability to
compute ground and excited state properties of clusters is also
important for the interpretation of diffraction experiments on small
$^4$He clusters.\,\cite{Sch.Toennies.95} Such experiments can be done
with transmission diffraction gratings, which have become available in
recent years.  The early experiments showed the presence of $^4$He
dimers, and other small clusters, but recently, more detailed physical
properties such as energies and bond lengths have been extracted from
these experiments.\,\cite{GriSchToeHeKoSt00} In spite of this
experimental progress, thus far, no computational method exists to
answer even the minimalist question of how many excited states small
$^4$He clusters have.  At a cluster conference in Germany in 1997,
this question made the list of pressing issues.

Another important issue that can be addressed by studying clusters is
the magnitude of three-body contributions to the interatomic potential
energy.  Evidence obtained by means of quantum Monte Carlo methods in
van der Waals complexes such as Ar$_n$HF and Ar$_n$HF dates back to at
least the mid nineties.\,\cite{Lewerenz96,NiyazBacicMoskowitzSchmidt96}
Hutson {\it et al.}\,\cite{HutsonLiuMoskovitzBacic99} studied the role
of three-body forces in more detail by a recent Discrete Variable
Representation (DVR) study of Ar$_n$HF with $n=2,3,4$, and work in
this field is still continuing.\,\cite{ToenniesVilesovWhaley}

Modifying diffusion Monte Carlo to calculate \keyword{vibrational
  states} is generally considered to be a difficult problem.  In
special cases, the excited state is the lowest energy state of a
particular symmetry, and then the standard fixed node approximation is
applicable.  Indeed, this approach has been used to compute tunneling
splittings for comparison with experiments on some water
clusters.\,\cite{GregoryClary95} For electronic excited states, in
many cases, approximate wave functions have nodal surfaces that are
sufficiently accurate that the fixed node diffusion Monte Carlo method
yields good energies but for many vibrational problems adequate nodal
surfaces are not available.

A method with the potential of addressing the excited state problem is
correlation function Monte Carlo.\,\cite{CeperleyBernu88,moreCFMC} A
promising new alternative diffusion Monte Carlo method for calculating
excited states employs the so-called projection operator imaginary
time spectral evolution (POITSE) approach.\,\cite{BlumeLewerenzWhaley97}
An important feature of both of these methods is that approximate
knowledge of the excited states can be built in from the start, and
the statistical accuracy of the estimates can be improved dramatically
in this way.  In fact, only rarely can one obtain results of
sufficient accuracy without these initial approximations.  However,
generating them can be quite difficult, and this is the problem we
address in this paper by means of Monte Carlo wave function
optimization.

More specifically, we address the problem of computing energies of
vibrationally \keyword{excited state}s by means of \qMC\ methods.  We
propose a method consisting of a combination of linear and non-linear
optimization schedules to generate optimized trial functions which are
used in a correlation function Monte Carlo computation.  The method is
applied to bosonic \keyword{van der Waals cluster}s.  As do other
\qMC\ methods, our approach has the ability to deal with systems with
strong anharmonicities and strong \qmechal\ fluctuations, which cause
the failure of conventional variational, normal mode, and basis set
methods.  In contrast to the fixed node diffusion \MC\ method, the one
discussed here does not require a priori knowledge of nodal surfaces.

The method\,\cite{NMA00} discussed in this paper relies on the use of
\keyword{optimized trial function}s for the excited states, and the
optimization method is explained in detail.  In applications, once the
optimized trial functions have been constructed, we use
\keyword{correlation function \MC}\ to reduce systematically the
variational bias of the energy estimates.  In principle, the imaginary
time spectral evolution (\keyword{POITSE}) method,\,\cite{BWjcp00}
(also see the paper by Whaley in this volume) could be used with the
optimized \keyword{excited state} wave functions we discuss here.  It
would be interesting to compare the relative merits of these two
\keyword{projection method}s.

\section{One state}
\label{section.one}
We consider \keyword{cluster}s of atoms of mass $\mu$, interacting
pairwise via a Lennard-Jones potential with core radius $\sigma$ and
well depth $\epsilon$.  In dimensionless form, the reduced pair
potential can be written as $v(r)=r^{-12}-2r^{-6}$ and the reduced
Hamiltonian as $\H=P^2/2m+V$, where $P^2/2m$ and $V$ are the total
reduced kinetic and potential energy operators. The only parameter is
the {\it dimensionless} inverse mass $m^{-1}=\hbar^2 /\mu \sigma^2
\epsilon$, which is proportional to the square of the \keyword{de Boer
  parameter},\,\cite{deBoer} a dimensionless measure of the relative
strength of quantum fluctuations.

We use the position representation, and denote by $R$ the $3\Nc$
Cartesian coordinates of the $\Nc$ atoms in the cluster. Suppose we
have a real-valued trial function $\psiT(R)$.  Typically, this trial
function may have 50-100 parameters and it may depend non-linearly on
these parameters.  First we recall how this wave function can be
optimized by minimization of the variance of the {\it local energy}
$\E(R)$, which is defined by
\begin{equation}
\H \psiT(R) \equiv \E(R) \psiT(R).
\end{equation}
Following Umrigar \etal,\,\cite{CyrusOptimization} one can minimize the variance
\begin{equation}
\chi^2 = \langle (\H-\langle\H\rangle)^2 \rangle,
\end{equation}
which in the position representation can be written as the variance of
the \keyword{local energy}.  Note that $\chi^2$ is nothing but the
square of the uncertainty in the energy, so that $\chi^2$ vanishes for
{\it any} eigenstate of the Hamiltonian $\H$.

The minimization of $\chi^2$ can be done by means of a \MC\ procedure
with the following steps:
\begin{enumerate}
\item Select a sample of configurations $R_1,\dots,R_s$ from the
  probability density $\psig^2$ (to be defined).
\item Evaluate:
\begin{eqnarray}
\Bm = \left(
\begin{array}{c}
\psiTh(R_1)\\
\vdots\\
\psiTh(R_s)
\end{array}
\right) 
\mbox{ and }
\Bm'=\left(
\begin{array}{c}
\psiTh\,'(R_1)\\
\vdots\\
\psiTh\,'(R_s)
\end{array}
\right),
\end{eqnarray}
where
\begin{equation}
\psiTh(R)={\psiT(R)\over\psig(R)}\mbox{ and }\psiTh\,'(R)={\H\psiT(R)\over\psig(R)}
\end{equation}
\item Find $\overline{\E}$ from least-squares solution of
\begin{equation}
\psiTh\,'(R_\sigma) = \overline{\E} \psiTh(R_\sigma),
\end{equation}
for $\sigma = 1,\dots,s$:
\begin{equation}
\overline{\E}={\displaystyle{\sum_{\sigma=1}^s \psiTh(R_\sigma)\psiTh\,'(R_\sigma)}
\over
{\displaystyle{\sum_{\sigma=1}^s \psiTh(R_\sigma)^2}}}.
\end{equation}
\item Vary the parameters in the trial function to minimize $\chi^2$,
  the normalized sum of squared residues defined by the previous step:
\begin{equation}
\chi^2={\displaystyle{\sum_{\sigma=1}^s \left[\psiTh\,'(R_\sigma) - 
\overline{\E} \psiTh(R_\sigma)\right]^2}
\over
\displaystyle{\sum_{\sigma=1}^s \psiTh(R_\sigma)^2}}.
\label{eq.variance}
\end{equation}
\end{enumerate}

The least-squares minimization of $\chi^2$ is done by means of the
Levenberg-Marquardt algorithm.\,\cite{quench} It should be noted that
this wave function optimization algorithm can, in principle be applied
to {\it any} eigenstate, but the basin of attraction of the ground
state typically is vastly bigger than that of any other eigenstate, so
that, the algorithm will practically never produce anything but an
approximation to the ground state, unless one starts the optimization
with a carefully designed initial wave function, such as, e.g.,
described in detail in the next section,.

For the purpose of optimizing only the ground state, the best choice
for the guiding function $\psig$, which is used to generate the sample
of configurations, is the optimized ground state wave function itself.
Since this function is only known at the end of the optimization, one
uses a reasonable initial approximation, if available.  Otherwise, a
few bootstrap iterations may be required.

For optimization of \keyword{excited state}s, one can use a power of
the optimized ground state trial wave function.  We have used a power
which is roughly in the range from one half to one third.  This has
the effect of increasing the range of configurations sampled with
appreciable probability.  The goal is to produce a sample that has
considerable overlap with the all excited states of interest.

\section{Several states}
\label{section.several}

Next we consider the problem of finding the `best' linear combination
of a number of given {\it elementary} \keyword{basis function}s
$\beta_1,\dots,\beta_n$.  Before we continue, we should explain our
terminology, since it reflects the procedure that will be used.  We
form linear combinations of the {\it \keyword{elementary basis
    function}s}.  These linear combinations depend on any non-linear
parameters that appear in the elementary basis functions; the linear
combinations will be optimized with respect to the non-linear
parameters by means of the general non-linear \keyword{trial function
  optimization} procedure described previously in Section
\ref{section.one}.  (See the discussion after
Eq.~(\ref{eq.psi.approx}) for a more explicit description of the
combined optimization procedure.)  Finally, these optimized basis
functions which serve as the {\it basis functions} in a \cfMC\ 
calculation,\,\cite{CeperleyBernu88,moreCFMC} and we shall return to
this in more detail later.

If the `best' linear combinations of elementary \keyword{basis
  function}s are defined in the sense that for such linear
combinations the expectation value of the energy is stationary with
respect to variation of the linear coefficients, the solution to this
problem is well known.  Being a linear problem, the solution requires
for its implementation traditional linear algebra.\,\cite{MacDonald}
The featured matrices consist of the matrix of overlap integrals of
the elementary basis functions, and the matrix of the Hamiltonian
sandwiched between them.  The trouble, of course, is that the required
matrix elements can be estimated only by means of \MC\ methods and the
\keyword{elementary basis function}s we employ for the
\keyword{cluster} problem.

Stationarity of the energy is equivalent to the least-squares
principle that is used in the following algorithm.  The latter can be
used with a very small sample of configurations, but in the limit of
an infinite sample it produces precisely the solution for which the
energy is stationary.

To find the optimal linear coefficients perform the following steps:
\begin{enumerate}
\item Select a sample of configurations $R_1,\dots,R_s$ from the
  probability density $\psig^2$ (as discussed previously).
\item Evaluate:
\begin{eqnarray}
\Bm &=&
\left(
\begin{array}{ccc}
\betah_1(R_1) & \cdots & \betah_n(R_1)\\
\vdots       & \vdots & \vdots\\
\betah_1(R_s) & \cdots & \betah_n(R_s)
\end{array}
\right),
\end{eqnarray}
and
\begin{eqnarray}
\Bm' &=&
\left(
\begin{array}{ccc}
\betah\,'_1(R_1) & \cdots & \betah\,'_n(R_1)\\
\vdots         & \vdots & \vdots\\
\betah\,'_1(R_s) & \cdots & \betah\,'_n(R_s)
\end{array}
\right),
\end{eqnarray}
where
\begin{equation}
\betah_i(R)={\beta_i(R)\over \psig(R)}\mbox{ and }\betah\,'_{i}(R)={\H\beta_i(R)\over \psig(R)}
\end{equation}
\item Find 
\begin{equation}
\Em=(\overline\E_{ij})_{i,j=1}^n
\label{eq.Em}
\end{equation}
from least-squares fit to
\begin{equation}
\betah\,'_i(R_\sigma) = \sum_{j=1}^n \betah_j(R_\sigma) \overline{\E}_{ji},
\label{eq.Hb}
\end{equation}
for $\sigma = 1,\dots,s$ and $i=1,\dots,n$.
\item Find the eigensystem of $\Em$ and write
\begin{equation}
\overline{\E}_{ij}=\sum_{k=1}^n \d i k \tE_k \,\dh j k,
\end{equation}
where the $\dh i k$ and  $\d j k$ are components of left and right
eigenvectors of $\Em$ with eigenvalue $\tE_k$.
\end{enumerate}
This algorithm yields an approximate expression for the eigenstate for
energy $E_k$:
\begin{equation}
\psi^{(k)}(R) \approx \tpsi k (R)= \sum_i \beta_i(R)\,\d i k.
\label{eq.psi.approx}
\end{equation}

In addition to an approximate eigenstate, this yields an eigenvalue
estimate which satisfies the following approximate
inequality\,\cite{MacDonald} for the excited state energy
\begin{equation}
E_k \alt \tE_k.
\end{equation}
The inequality holds rigorously in the absence of statistical noise,
i.e., if an infinite \MC\ sample is used, or, if by other means, the
\qmechal\ overlap integrals and matrix elements, corresponding to the
matrices $\Nm$ and $\Hm$ defined in Eq.~(\ref{eq.Em.sol}), are
evaluated exactly.

Before we discuss some technical details of the linear optimization,
let us summarize the full optimization scheme for excited state $k$.
We iteratively minimize the variance of the local energy of the
approximate eigenstate given in Eq.~(\ref{eq.psi.approx}), where the
variance is defined in Eq.~(\ref{eq.variance}).  This minimization is
with respect to the non-linear parameters that appear in the
elementary basis functions as to be defined exlicitly in
Section~\ref{sec.elementary}.  During the least-squares minimization,
for any given choice of non-linear parameters, the linear parameters
$\d i k$ are defined by the algorithm given in this section, which indeed
produces the state given by Eq.~(\ref{eq.psi.approx}) required in the
definition of the variance in Eq.~(\ref{eq.variance}).

In the ideal case that the \keyword{basis function}s are linear
combinations of no more than $n$ true eigenfunctions of the
Hamiltonian, the previous algorithm yields the true eigenvalues, even
for a finite \MC\ sample, unless it fails altogether for lack of
sufficient independent data.

To find the least-squares solution for $\Em$ from Eq.~(\ref{eq.Hb})
we write the latter in the form
\begin{equation}
\Bm'=\Bm\Em.
\end{equation}
Multiply through from the left by $\Bm^T$, the transpose of $\Bm$, and
invert to obtain
\begin{equation}
\Em=(\Bm^T\Bm)^{-1}(\Bm^T\Bm')\equiv\Nm^{-1}\Hm.
\label{eq.Em.sol}
\end{equation}
It is simple to verify that indeed this yields the least-squares
solution of Eqs.~(\ref{eq.Hb}).  Note that the matrix $\Hm$ becomes
symmetric only in the limit of an infinite sample; symmetrizing it for a
finite sample destroys the zero-variance propoerty of
Eq.~\ref{eq.Em.sol}.

It is is well-known that the solution for $\Em$ as written in
Eq.~(\ref{eq.Em.sol}) is numerically
unstable.\,\cite{NumericalRecipes.SVD} This is a consequence of the fact
that the matrix $N$ is ill conditioned if the $\beta_k$ are nearly
linearly dependent, which indeed is the case for our elementary basis
functions.  The solution to this problem is to use a singular value
decomposition to obtain a numerically defined and regularized inverse
$\Bm^{-1}$.  In terms of the latter, one finds from
Eq.~(\ref{eq.Em.sol})
\begin{equation}
\Em=\Bm^{-1}\Bm'.
\end{equation}

More explicitly, one uses a singular value decomposition to
write\,\cite{GolubVanLoan}
\begin{equation}
\Bm=\Um \Sm \tr \Vm,
\label{eq.svd}
\end{equation}
where $\Um$ and $\Vm$ are square orthogonal matrices respectively of
order $s$, the sample size, and $n$, the number of 
\keyword{elementary basis function}s, while $S_r$ is a rectangular $s \times n$ matrix
with zeroes everywhere except for its leading diagonal elements
$\sigma_1 \ge \sigma_2 \ge \sigma_r > 0$; $r$ is chosen such that the
remaining singular values are sufficiently close to zero to be
ignored.  In our applications, we ignored all singular values
$\sigma_k$ with $\sigma_k < 10^3 \sigma_1 \epsilon_{\rm dbl}$, where
$\epsilon_{\rm{dbl}}$ is the double precision machine accuracy.  This
seems a reasonable choice, but we have no compelling argument to
justify it.

From Eq.~(\ref{eq.svd}) one obtains
\begin{equation}
\Em=\Vm_r \Sm_r^{-1} \tr \Um_r \Bm',
\label{eq.Esvd}
\end{equation}
where $\Um_r$ is the $s \times r$ matrix consisting of the first $r$
columns of $\Um$; $\Vm_r$ is the $n\times r$ matrix likewise obtained
from $\Vm$; and $\Sm_r$ is the $r\times r$ upper left corner of $\Sm$.

\section{Elementary basis functions}
\label{sec.elementary}
We used \keyword{elementary basis function}s of the general form
introduced in Ref.~\,\onlinecite{MN94}. Rotation and translation symmetry
are built into these functions by writing them as functions of all
interparticle distances.  First of all, we introduce a scaling
function with values that change appreciably only in the range of
interparticle distances that occur in the \keyword{cluster} configurations with
appreciable probability.  For this purpose we first introduce a
piecewise linear function $f$.  This function has three parameters:
$x_1<x_2<x_3$, which define the four linear segments of the continuous
function $f$:
\begin{equation}
f(x)=\left\{\begin{array}{r}
-1 \mbox{ for } x\le x_1,\\
 0 \mbox{ for } x=x_2,\\
 1 \mbox{ for } x\ge x_3.

\end{array}\right.
\end{equation}
The parameters are determined by the relevant length scales of the
system.  The parameter $x_1$ sets the scale for how close two atoms
can get with reasonable probability; $x_2$ roughly equals the most
likely interparticle distance; and $x_3$ is the distance at which one
expects the onset of the long distance asymptotic regime.  Possibly,
one could drop $x_2$ and use a simpler function consisting of three
linear segments only.

The function $f$ has no continuous derivatives and cannot be used
directly as a scaling function.  Instead, we use the generalized
Gaussian transform
\begin{equation}
\hat f(x)=\int_{-\infty}^{\infty} f(x') \exp-{(x-x')^2\over 2 c x}\,dx',
\end{equation}
with $c=0.1$.

In their most general form the wave functions in Ref.\,\onlinecite{MN94}
contain five-body correlations, but in the work reported here we have
only used \keyword{three-body correlations} and for completeness we shall
describe the construction of these functions explicitly.

Choose three of the $\Nc$ atoms.  Suppose they have labels
$\alpha,\beta,$ and $\gamma$ and Cartesian coordinates
$\r_\alpha,\r_\beta$ and $\r_\gamma$.  This defines three scaled
interatomic distances
\begin{equation}
\left.
\begin{array}{c}
\rh_\alpha=\fh(|\r_\beta-\r_\gamma|)\\
\rh_\beta=\fh(|\r_\gamma-\r_\alpha|)\\
\rh_\gamma=\fh(|\r_\alpha-\r_\beta|)
\end{array}\right\}.
\end{equation}
Define three \keyword{invariants} as sums of powers of these variables
\begin{equation}
I_p=\rh_\alpha^{\,\,p}+\rh_\beta^{\,\,p}+\rh_\gamma^{\,\,p},
\label{eq.invariants}
\end{equation}
with $p=1,2,3$. Clearly, any polynomial in the invariants $I_1,I_2$
and $I_3$ is symmetric with respect to permutation of the labels
$\alpha,\beta,$ and $\gamma$.  A convenient property of these variables
is that the reverse is also true: any symmetric polynomial in the
three scaled distances can be written as a polynomial in the invariants
$I_1,I_2$ and $I_3$.  This makes it simple to parameterize these
symmetric polynomials.

In terms of the \keyword{invariants} we define `minimal polynomials'
$s_i$ as follow: pick a monomial in $I_1,I_2,$ and $I_3$ and sum over
all possible ways of choosing three atoms $\alpha,\beta,$ and
$\gamma$.  These polynomials are minimal in the sense that one cannot
omit any single term without violating the bosonic symmetry.

In addition to \keyword{bosonic symmetry}, we impose short and
long-distance boundary conditions.  This yields the following form for
the \keyword{elementary basis function}s
\begin{equation}
\beta_i(R)=s_i(R)
\exp\left({\sum_j a_j\,s_j(R)}
-\sum_{\sigma<\tau} \left(\kappa_k\, r_{\sigma\tau}+
{\sqrt m\over 5\,r_{\sigma\tau}^5}\right)\right)
\label{eq.elbasis}
\end{equation}
with
\begin{equation}
\kappa_k={2 \over \Nc-1}\sqrt{-m\tE_k\over \Nc}.
\end{equation}
As discussed in detail in Ref.\,\onlinecite{MN94}, the
$r_{\sigma\tau}^{-5}$ term in the exponent and its coefficient are
chosen so that, when two atoms approach each other, the strongest
divergence in the local energy, i.e. the Lennard-Jones
$r_{\sigma\tau}^{-12}$ divergence, is canceled by the divergence in
the local kinetic energy.  The energy $\tE_k$ is determined
self-consistently by iteration; one or two iterations typically
suffice.  The specific form of the decay constant is chosen on the
basis of two assumptions.  The energy is assumed to be proportional to
the number of atom pairs in the \keyword{cluster}.\,\cite{LDW91} This
is reasonable for small clusters, but for larger ones this should
probably be modified to reflect the expectation that the energy is
proportional to the average number of nearest neighbor pairs.  The
second assumption is that if one atom is far away from all others, the
wave function can be written as the product of an $\Nc-1$ cluster wave
function and an exponentially decaying part that carries a fraction of
the total energy equal to the number of bonds connecting that atom to
the others.

The $a_j$ in Eq.~(\ref{eq.elbasis}) are non-linear variational
parameters.  Their optimal values are re-optimized for each
\keyword{excited state}.  In principle, one could optimize all
non-linear parameters, including those that appear in the scaling
function and the factors that impose the boundary conditions.
However, it has been our experience that this produces strongly
correlated variational parameters and results in unstable fits.

\section{Reduction of variational errors}

The linear and non-linear optimization procedures described above are
used to generate \keyword{basis function}s for a \cfMC\ 
calculation,\,\cite{CepBer88} which increases the statistical accuracy
of the energy estimates and reduces the systematic errors due to
imperfections of the variational functions.  The number of these
\keyword{basis function}s is much smaller than the number of
\keyword{elementary basis function}s that appear in the linear
combinations.  The advantage of not using {\it all}
\keyword{elementary basis function}s for \cfMC\ purposes can be
understood as follows.

Suppose that the optimization phase yields states $\ket{\psiT^{(k)}}$
with \hbox{$k = 1,\dots,$} \hbox{$n'<n$.}  Correlation function \MC\ 
in a statistical sense yields the \keyword{basis function}s
\begin{equation}
\ket{\psiT^{(i)}(t)}\equiv e^{\displaystyle{-\H t}}\ket{\psiT^{(i)}}.
\end{equation}
As $t$ increases, the spectral weight of undesirable \keyword{excited
  state}s, i.e., states $k$ with $E_k>E_{n'}$ is decreased.  That is
desirable, but at the same time all \keyword{basis function}s approach
the ground state and therefore become more nearly linearly dependent.
More explicitly, one has \MC\ estimates of the following the
generalization of Eq.~(\ref{eq.Em.sol})
\begin{equation}
\Em(t)=\Nm(t)^{-1}\Hm(t),
\label{eq.Emt}
\end{equation}
with
\begin{equation}
N_{ij}(t)=\br {\psiT^{(i)}(t)}\ket{\psiT^{(j)}(t)}
\end{equation}
and
\begin{equation}
H_{ij}(t)=\bra {\psiT^{(i)}(t)}\H \ket{\psiT^{(j)}(t)}.
\end{equation}
Again, trouble is caused by an ill-conditioned matrix, which in this
case is $\Nm(t)$, and increasingly so for increasing values of the
projection time $t$.  Obviously, the better are the trial states
$\ket{\psiT^{(i)}}$ and the fewer is their number, the less severe is
this problem.  We should also point out in this context that the
singular value decomposition cannot be used in this case.  The reason
is that the analogs of the matrices $\Bm$ and $\Bm'$ become too big to
store for \MC\ samples of the size required in the \cfMC\ runs.

\section{Results}

We computed reduced energies for clusters consisting of various noble
gas atoms He, Ne, Ar, and Kr, corresponding respectively to the
dimensionless inverse masses $m^{-1}=9.61\times 10^{-3},\ 7.092\times
10^{-3},\ 6.9\ 635\times 10^{-4}$ and $1.9\ 128\times 10^{-4}$.
Quantities with the dimension of energy can be reconstructed from the
following values for the corresponding well depths $\epsilon/k_{\rm
  B}\rm{K}=10.22$, $35.6$, $119.4$, $164.0$, $222.3$.

In Table \ref{Trimer.table} we compare results obtained with our \MC\ 
method with results of Leitner \etal,\,\cite{LDW91} which were
obtained by the discrete variable representation (DVR) method.  With
the exception of the fifth state of Ne, the \MC\ results agree with or
improve the DVR results.  In some cases, the disagreement can be
attributed to lack of convergence of the DVR
results.\,\cite{LeitnerPrivate} Similar effects were observed in prior
variational calculations, which employed a precursor of the
optimization scheme discussed here.\,\cite{MNijmp00} The discrepancy
for the fifth state of Ne may be an illustration of a weakness of the
\cfMC\ method, as it is commonly implemented, namely the difficulty of
estimating the statistical and systematic errors.

As is well known, in \dMC\ computations one has to contend with
time-step errors, due to the short-time approximation that has to be
used for the imaginary time evolution operator $\exp -t\H$. The same
applies to the \cfMC\ method.  For all results reported here, we
repeated the calculations for a range of time steps to verify that the
time-step errors were smaller than the statistical and other
systematic errors. See Ref.~\,\onlinecite{NMA00} for further details.

There can be problems both with obtaining reliable estimates of the
statistical errors and with establishing convergence as a function of
projection time $t$ [\cf\ Eq.~(\ref{eq.Emt})].  This is a consequence
of the fact that the data for different values of the projection time
are strongly correlated since they are obtained from the same \MC\ 
data.  Correlated noise may introduce false trends or obscure true
ones, a problem that in principle can be solved by performing
independent runs for different projection times, but that would
greatly increase the computation time.

Unreliable statistical error estimates may also come about because the
\cfMC\ calculation takes the form of a
pure-\dMC\,\cite{caffarel,UNR93} calculation.  The algorithm used for
the latter features weights consisting of a number of fluctuating
factors proportional to the projection time $t$.  Consequently, as the
projection time $t$ increases, the variance of the estimators
increases and they acquire a significantly non-Gaussian
distribution,\,\cite{Hetherington} which renders error bars computed
in the standard way increasingly misleading.  Conceivably, one could
reduce the severity of this effect by using branching random
walks,\,\cite{nighNATO} as is done in standard \dMC, or by means of
reptation \MC.\,\cite{baroni-moroniNATO}

In Table \ref{Argon.table} we present results for the energies of the
first five levels of Ar \keyword{cluster}s of sizes four through seven. Our
method allows one to go beyond seven atom clusters, but, as one can
see from Table \ref{Argon.table}, the statistical errors increase with
system size.  To obtain more accurate results for larger clusters it
would probably be helpful to include higher order correlations in the
wave function, since the degrees of the polynomials were chosen
sufficiently high that increasing them further no longer improves the
quality of the trial functions.

\begin{figure}[htb]
\begin{center}
\setlength{\unitlength}{0.240900pt}
\ifx\plotpoint\undefined\newsavebox{\plotpoint}\fi
\sbox{\plotpoint}{\rule[-0.200pt]{0.400pt}{0.400pt}}%
\begin{picture}(1297,1080)(0,0)
\font\gnuplot=cmr10 at 10pt
\gnuplot
\sbox{\plotpoint}{\rule[-0.200pt]{0.400pt}{0.400pt}}%
\put(201.0,163.0){\rule[-0.200pt]{4.818pt}{0.400pt}}
\put(181,163){\makebox(0,0)[r]{-2.5}}
\put(1257.0,163.0){\rule[-0.200pt]{4.818pt}{0.400pt}}
\put(201.0,338.0){\rule[-0.200pt]{4.818pt}{0.400pt}}
\put(181,338){\makebox(0,0)[r]{-2}}
\put(1257.0,338.0){\rule[-0.200pt]{4.818pt}{0.400pt}}
\put(201.0,513.0){\rule[-0.200pt]{4.818pt}{0.400pt}}
\put(181,513){\makebox(0,0)[r]{-1.5}}
\put(1257.0,513.0){\rule[-0.200pt]{4.818pt}{0.400pt}}
\put(201.0,689.0){\rule[-0.200pt]{4.818pt}{0.400pt}}
\put(181,689){\makebox(0,0)[r]{-1}}
\put(1257.0,689.0){\rule[-0.200pt]{4.818pt}{0.400pt}}
\put(201.0,864.0){\rule[-0.200pt]{4.818pt}{0.400pt}}
\put(181,864){\makebox(0,0)[r]{-0.5}}
\put(1257.0,864.0){\rule[-0.200pt]{4.818pt}{0.400pt}}
\put(201.0,1039.0){\rule[-0.200pt]{4.818pt}{0.400pt}}
\put(181,1039){\makebox(0,0)[r]{0}}
\put(1257.0,1039.0){\rule[-0.200pt]{4.818pt}{0.400pt}}
\put(201.0,163.0){\rule[-0.200pt]{0.400pt}{4.818pt}}
\put(201,122){\makebox(0,0){0}}
\put(201.0,1019.0){\rule[-0.200pt]{0.400pt}{4.818pt}}
\put(355.0,163.0){\rule[-0.200pt]{0.400pt}{4.818pt}}
\put(355,122){\makebox(0,0){0.05}}
\put(355.0,1019.0){\rule[-0.200pt]{0.400pt}{4.818pt}}
\put(508.0,163.0){\rule[-0.200pt]{0.400pt}{4.818pt}}
\put(508,122){\makebox(0,0){0.1}}
\put(508.0,1019.0){\rule[-0.200pt]{0.400pt}{4.818pt}}
\put(662.0,163.0){\rule[-0.200pt]{0.400pt}{4.818pt}}
\put(662,122){\makebox(0,0){0.15}}
\put(662.0,1019.0){\rule[-0.200pt]{0.400pt}{4.818pt}}
\put(816.0,163.0){\rule[-0.200pt]{0.400pt}{4.818pt}}
\put(816,122){\makebox(0,0){0.2}}
\put(816.0,1019.0){\rule[-0.200pt]{0.400pt}{4.818pt}}
\put(970.0,163.0){\rule[-0.200pt]{0.400pt}{4.818pt}}
\put(970,122){\makebox(0,0){0.25}}
\put(970.0,1019.0){\rule[-0.200pt]{0.400pt}{4.818pt}}
\put(1123.0,163.0){\rule[-0.200pt]{0.400pt}{4.818pt}}
\put(1123,122){\makebox(0,0){0.3}}
\put(1123.0,1019.0){\rule[-0.200pt]{0.400pt}{4.818pt}}
\put(1277.0,163.0){\rule[-0.200pt]{0.400pt}{4.818pt}}
\put(1277,122){\makebox(0,0){0.35}}
\put(1277.0,1019.0){\rule[-0.200pt]{0.400pt}{4.818pt}}
\put(201.0,163.0){\rule[-0.200pt]{259.208pt}{0.400pt}}
\put(1277.0,163.0){\rule[-0.200pt]{0.400pt}{211.028pt}}
\put(201.0,1039.0){\rule[-0.200pt]{259.208pt}{0.400pt}}
\put(41,601){\makebox(0,0){$-|E_k|^{1 \over 2}$}}
\put(739,61){\makebox(0,0){$m^{-{1 \over 2}}$}}
\put(243,58){\makebox(0,0){Kr}}
\put(282,233){\makebox(0,0){Ar}}
\put(460,233){\makebox(0,0){Ne}}
\put(1154,233){\makebox(0,0){He}}
\put(201.0,163.0){\rule[-0.200pt]{0.400pt}{211.028pt}}
\put(232,181){\vector(-1,0){31}}
\put(243,128){\vector(0,1){35}}
\put(282,198){\vector(0,-1){35}}
\put(460,198){\vector(0,-1){35}}
\put(1154,198){\vector(0,-1){35}}
\put(488,969){\makebox(0,0)[r]{$k=1: $}}
\put(324,280){\raisebox{-.8pt}{\makebox(0,0){$\Diamond$}}}
\put(418,354){\raisebox{-.8pt}{\makebox(0,0){$\Diamond$}}}
\put(508,423){\raisebox{-.8pt}{\makebox(0,0){$\Diamond$}}}
\put(636,521){\raisebox{-.8pt}{\makebox(0,0){$\Diamond$}}}
\put(733,592){\raisebox{-.8pt}{\makebox(0,0){$\Diamond$}}}
\put(816,650){\raisebox{-.8pt}{\makebox(0,0){$\Diamond$}}}
\put(888,699){\raisebox{-.8pt}{\makebox(0,0){$\Diamond$}}}
\put(954,741){\raisebox{-.8pt}{\makebox(0,0){$\Diamond$}}}
\put(1014,778){\raisebox{-.8pt}{\makebox(0,0){$\Diamond$}}}
\put(1071,811){\raisebox{-.8pt}{\makebox(0,0){$\Diamond$}}}
\put(1123,840){\raisebox{-.8pt}{\makebox(0,0){$\Diamond$}}}
\put(1173,868){\raisebox{-.8pt}{\makebox(0,0){$\Diamond$}}}
\put(558,969){\raisebox{-.8pt}{\makebox(0,0){$\Diamond$}}}
\put(488,928){\makebox(0,0)[r]{$k=2: $}}
\put(324,314){\makebox(0,0){$+$}}
\put(418,401){\makebox(0,0){$+$}}
\put(508,470){\makebox(0,0){$+$}}
\put(636,569){\makebox(0,0){$+$}}
\put(733,635){\makebox(0,0){$+$}}
\put(816,696){\makebox(0,0){$+$}}
\put(888,750){\makebox(0,0){$+$}}
\put(954,801){\makebox(0,0){$+$}}
\put(1014,843){\makebox(0,0){$+$}}
\put(1071,871){\makebox(0,0){$+$}}
\put(1123,915){\makebox(0,0){$+$}}
\put(558,928){\makebox(0,0){$+$}}
\sbox{\plotpoint}{\rule[-0.400pt]{0.800pt}{0.800pt}}%
\put(488,887){\makebox(0,0)[r]{$k=3: $}}
\put(324,325){\raisebox{-.8pt}{\makebox(0,0){$\Box$}}}
\put(418,418){\raisebox{-.8pt}{\makebox(0,0){$\Box$}}}
\put(508,485){\raisebox{-.8pt}{\makebox(0,0){$\Box$}}}
\put(636,600){\raisebox{-.8pt}{\makebox(0,0){$\Box$}}}
\put(733,690){\raisebox{-.8pt}{\makebox(0,0){$\Box$}}}
\put(816,760){\raisebox{-.8pt}{\makebox(0,0){$\Box$}}}
\put(888,809){\raisebox{-.8pt}{\makebox(0,0){$\Box$}}}
\put(954,835){\raisebox{-.8pt}{\makebox(0,0){$\Box$}}}
\put(1014,871){\raisebox{-.8pt}{\makebox(0,0){$\Box$}}}
\put(558,887){\raisebox{-.8pt}{\makebox(0,0){$\Box$}}}
\end{picture}
\end{center}
\caption{$-\sqrt{-E_k}$ for lowest three vibrational levels ($k=1,2,3$) of four
  particle \keyword{cluster}s vs $m^{-{1\over 2}}$.  The estimated errors for
  most energies are smaller, than the plot symbols.  Results for level
  $k=3$ become unreliable near He and have not been included. The
  vertical arrows indicate Kr, Ar, Ne, and He; the horizontal arrow
  indicates the classical value -$\sqrt 6$.}
\label{fig.Evsm}
\end{figure}
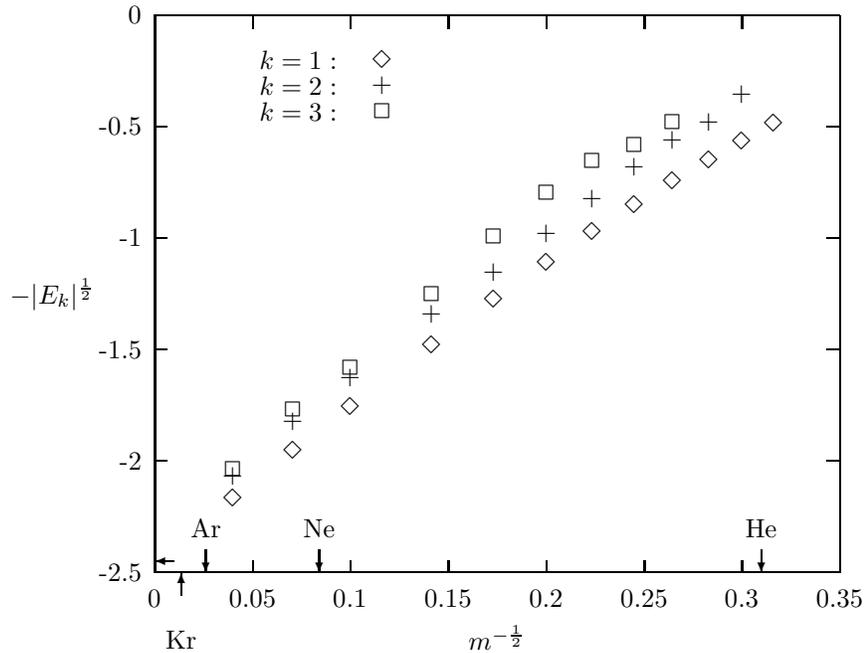

Figure~\ref{fig.Evsm} contains three energy levels as a function of
mass for four particle \keyword{clusters}.  The harmonic approximation
implies that for large masses the energy is a linear function of
$m^{-{1\over 2}}$.  We expect the energy to vanish quadratically in
the vicinity of the dissociation limit.  The results are therefore
plotted using variables that yield linear dependence both for large
masses and for energies close to zero.\,\cite{MMN96} As the energy
levels approaches zero, both the optimization and the projection
methods begin to fail, and correspondingly data points are missing.

In the \keyword{elementary basis function}s, we typically used
polynomials of degree ten in the prefactors and of degree three in the
exponent.  The \dMC\ runs used on the order of a million steps with a
time step of a couple of tenths.  The longer runs typically took a few
hours on a four processor SGI Origin 200.

\begin{table}
\caption{
Comparisson with DVR results of reduced energy $E_k$ of vibrational
energy levels of noble gas \keyword{trimers}; the estimated errors 
(which reflect a combination of statistical and systematic errors, 
as explained in the text) are a few units in the least significant decimal.}
\vspace{5pt}
\begin{center}
\begin{tabular}{l|ll||ll}
$k$&\hskip 35pt Ne$_3$  &  &\hskip 43pt Ar$_3$     &   \\
\hline
&\hskip 15pt MC&\hskip 3pt DVR&\hskip 20pt MC&\hskip 3pt DVR\\
\hline
1& -1.719~560 &-1.718& -2.553~289~43&-2.553 \\
2& -1.222~83  &-1.220& -2.250~185~5 &-2.250 \\
3& -1.142~0   &-1.138& -2.126~361   &-2.126 \\
4& -1.038     &-1.035& -1.996~43    &-1.996 \\
5& -0.890     &-0.898& -1.946~7     &-1.947 \\
\end{tabular}
\end{center}
\label{Trimer.table}
\end{table}

\begin{table}
\caption{Reduced energy of vibrational levels $E_k$ of Ar \keyword{clusters};
the estimated  errors (which reflect a combination of statistical and systematic errors, 
as explained in the text) are a few units in the least significant decimal.}
\vspace{20pt}
\vspace{5pt}
\begin{center}
\begin{tabular}{l|l|l|l|l}
$k$&  Ar$_4$ & Ar$_5$ & Ar$_6$    & Ar$_7$\\
\hline
1& -5.118~11 & -7.785~1 & -10.887~9 & -14.191\\
2& -4.785    & -7.567   & -10.561   & -13.969\\
3& -4.674    & -7.501   & -10.51    & -13.80 \\
4& -4.530    & -7.39    & -10.46    & -13.74 \\
5& -4.39     & -7.36    & -10.35    & -13.71 \\
\end{tabular}
\end{center}
\label{Argon.table}
\end{table}

\section{Discussion}

We presented a scheme to optimize trial functions to be used in \qMC\ 
calculations for excited states.  We applied this scheme to
Lennard-Jones clusters, but it can also be applied to systems
interacting with more realistic pair- and even three-body potentials.
Only the elementary basis functions would have to be modified, since
it should be noted that the $r^{-5}$ contribution in the exponent in
Eq.~(\ref{eq.elbasis}) is specifically tailored to suppress the
$r^{-12}$ divergence of the local energy for the Lennard-Jones
potential.

The method can also be applied straightforwardly to clusters of
experimental interest containing chromophores.  In that case, more
drastic changes will have to be made to the elementary basis
functions.  E.g., the invariants defined in Eq.~(\ref{eq.invariants}),
which guarantee full bosonic symmetry, are of course no longer
appropriate.  In general, the optimization method as such can be used
for completely different, even fermionic systems, as long as the
elementary basis functions are adapted for such applications.

As far as the projection phase of the computations is concerned, the
standard implementation of \cfMC\ in the form of pure-\dMC\ will
become inefficient from systems of increasing size and might have to
be replaced by a branching algorithm based on branching random walks
or reptation \MC.\,\cite{baroni-moroniNATO} In this context also the
use of our optimization scheme in conjunction with the POITSE method
is an interesting possibility.\,\cite{BlumeLewerenzWhaley97}

The most serious problem with the results presented in this paper
shows up at small masses.  Clearly, this is caused by a deficiency in
the variational freedom of the elementary basis functions employed in
the currrent applications.  The incorporation of \keyword{four-body
  correlations}, which seems to be all that the current basis
functions are lacking, is likely to yield more accurate results for
atoms of small mass.

\section*{Acknowledgments}
This research was supported by the (US) National Science Foundation
(NSF) through Grant DMR-9725080.  It is our pleasure to thank David
Freeman and Cyrus Umrigar for valuable discussions.

\end{document}